\documentclass[11pt]{article}
\usepackage{moriond,epsfig}
\usepackage{wrapfig}
\usepackage{psfrag}
\usepackage{amsmath,graphicx}

\newcommand{\wh}[1]{\widehat{#1}}
\def\spa#1.#2{\left\langle#1\,#2\right\rangle}
\def\spb#1.#2{\left[#1\,#2\right]}

\def\AB#1#2#3{\langle#1|#2|#3]}

\def\A#1#2{\langle#1#2\rangle}
\def\B#1#2{[#1#2]}
\def\la{\langle}
\def\ra{\rangle}
\def\sst{\scriptscriptstyle}
\def\NP{N_{P}}
\def\NF{N_{F}}

\DeclareMathOperator{\tr}{ {\rm tr}}

\DeclareMathOperator{\Tri}{ {\rm F}^{1m}_3}
\DeclareMathOperator{\Ftme}{ {\rm F}^{2me}_4}
\DeclareMathOperator{\Fom}{ {\rm F}^{1m}_4}

\begin{document}
\begin{flushright}
IPPP/07/15\\ SACLAY-SPHT-T07/051\\May 2007
\end{flushright}
\vspace*{2cm}
\title{Higgs Amplitudes From Twistor Inspired Methods}
\author{
S. D. Badger$^{a}$, E.W.N. Glover$^{b}$ and Kasper Risager$^{c}$}
\address{$^{a}$ Service de Physique Theorique, CEA/Saclay, 91191 Gif-sur-Yvette, France,\\
$^{b}$ Department of Physics, University of Durham, Durham, DH1 3LE, UK,\\
$^{c}$ Niels Bohr Institute, Blegdamsvej 17, DK-2100, Copenhagen, Denmark}

\maketitle
\abstracts{
We illustrate the use of new on-shell methods, 4-dimensional unitarity cuts combined with 
on-shell recursions relations, 
by computing the $A_4^{(1)}(\phi,1^-,2^-,3^+,4^+)$ amplitude in the large top mass limit
where the Higgs boson couples to gluons through an effective interaction.
}


\section{Introduction}

The time for experiments at the LHC is approaching fast and it will be extremely important to have
accurate predictions of Standard Model processes if we hope to find signals of new physics. Recent
advances in on-shell techniques \cite{BDK:1lonshell,BDK:1lrecfin,Bern:bootstrap} have made it possible to
calculate compact analytic expressions for multiparticle scattering amplitudes at one loop. Here we
consider the application of these so called ``twistor" inspired methods to one loop amplitudes with
a massive, colourless scalar (the Higgs boson) two negative helicity gluons and 
two positive helicity gluons\footnote{Analytic expressions for the $\phi$-MHV amplitude
an arbitrary number of positive
helicity gluons can be found in~\cite{Badger:phimm}}. This builds on work for amplitudes with simpler helicity configurations
\cite{Berger:higgsrecfinite,Badger:1lhiggsallm}. 

We consider the leading colour contribution to the one-loop amplitudes and, following
\cite{Bern:bootstrap}, we split them into evaluating the ``pure''
4-dimensional cut-constructible ${C}_4$ and rational ${R}_4$.
\begin{align}
	A_4^{(1)}(\phi,1^-,2^-,3^+,4^+) &=  {C}_4(\phi,1^-,2^-,3^+,4^+)+ {R}_4(\phi,1^-,2^-,3^+,4^+).
\end{align}
The pure cut piece contains all (poly)logarithmic terms ($\log,{\rm Li}_2,\pi^2$) which can be found by
computing the unitarity cuts in 4 dimensions \cite{BDDK:uni1,BDDK:uni2}. The remaining rational terms
can then be evaluated using on-shell recursion relations.

\section{The Model}

In the Standard Model the Higgs boson couples to gluons through
a fermion loop where the dominant contribution is from the top quark.
It is well known that for large $m_t$, the top quark loop can be integrated out leading
to the effective interaction,
\begin{equation}
 {\cal L}_{\sst H}^{\rm int} =
  \frac{C}{2}\, H \tr G_{\mu\nu}\, G^{\mu\nu}  \ .
 \label{HGGeff}
 \end{equation}
In the Standard Model, and to leading order in $\alpha_s$,
the strength of the interaction is given by $C = \tfrac{\alpha_s}{6\pi v} + \mathcal{O}(\alpha_s^2)$,
with $v = 246$~GeV. 
$C$ has been calculated up to order ${\cal{O}}(\alpha_{s}^{4})$~\cite{Chetyrkin:heffalpha3}.

The MHV structure of the Higgs-plus-gluons amplitudes is best
elucidated \cite{Dixon:MHVhiggs} by considering $H$ to be the real
part of a complex field $\phi = \frac{1}{2}( H + i A )$, so that
\begin{eqnarray}
 {\cal L}^{\rm int}_{H,A} &=&
\frac{C}{2} \Bigl[ H \tr G_{\mu\nu}\, G^{\mu\nu}
             + i A \tr G_{\mu\nu}\, {}^*G^{\mu\nu} \Bigr]
 \label{effinta}\nonumber \\
&=&
C \Bigl[ \phi \tr G_{{\sst SD}\,\mu\nu}\, G_{\sst SD}^{\mu\nu}
 + \phi^\dagger \tr G_{{\sst ASD}\,\mu\nu} \,G_{\sst ASD}^{\mu\nu} \Bigr]
 \label{effintb}
\end{eqnarray}
The amplitudes of the $\phi$ and $\phi^\dagger$ turn out to be much simpler than the corresponding
$H$ and $A$ fields and so we proceed by calculating helicity amplitudes for gluons coupling to the
$\phi$ and then construct the $\phi^\dagger$ amplitudes using parity symmetry. The full Higgs
amplitudes are then made from the sum of $\phi$ and $\phi^\dagger$ amplitudes.

\section{Cut Constructible Contributions}

The unitarity method relies on sewing together tree-level amplitudes with on-shell propagators. Here
we use the method of Brandhuber, Spence and Travaglini \cite{Brandhuber:n4} which uses the off-shell
continuation used for the tree-level MHV rules \cite{Cachazo:MHVtree} to sew together tree MHV
amplitudes. The subsequent integration over the continuation parameter, $z$, reconstructs only the
specific parts of the integral functions which have cuts in the considered channel. The cut
integrals that are encountered have been considered previously by Van Neerven
\cite{vanNeerven:dimreg}.

\begin{figure}[h]
\begin{center}
	\psfrag{phi}{\tiny$\phi$}
	\psfrag{1}{\tiny$1-$}
	\psfrag{2}{\tiny$2-$}
	\psfrag{3}{\tiny$3+$}
	\psfrag{4}{\tiny$4+$}
	\includegraphics[width=12cm]{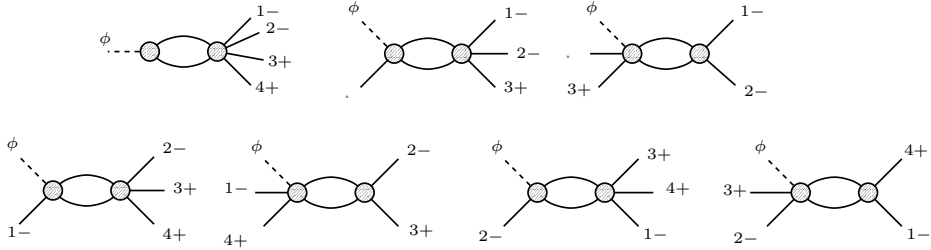}
	\caption{The topologies for the cut-constuctible part of the four gluon $\phi$-MHV amplitude.}
	\label{fig:cdiags}
\end{center}
\end{figure}

The
tree-level QCD amplitudes have been known for some time \cite{Parke:ngluon} and the relevant $n$-point
tree-level $\phi$-amplitudes have been recently computed using the CSW method
\cite{Dixon:MHVhiggs,Badger:MHVhiggs2}. The topologies obtained by 
sewing together all
possible configurations of joining a tree-level $\phi$-MHV amplitude with a pure QCD amplitude 
are shown in figure \ref{fig:cdiags}
where both gluons and fermions are allowed to circulate in the loop~\cite{Badger:phimm}. 
The final result is given by~\cite{Badger:phimm}:
\begin{align}
	C_4(\phi,1^-,&2^-,3^+,4^+) =c_\Gamma A^{(0)}_4(\phi,1^-,2^-,3^+,4^+)
	\Bigg[U_4\nonumber \\
	&+\bigg(
	\frac{\NP}{3}\frac{\langle1432\rangle^3}{\spa{1}.{2}^3}L_3(s_{341},s_{41})
	+\frac{\NP}{3}\frac{\langle2341\rangle^3}{\spa{2}.{1}^3}L_3(s_{234},s_{23})
	\nonumber \\
	&\phantom{\sum_{i=4}^{n}\bigg(}
	-\frac{\NP}{2}\frac{\langle1432\rangle^2}{\spa{1}.{2}^2}L_2(s_{341},s_{41})
	-\frac{\NP}{2}\frac{\langle2341\rangle^2}{\spa{2}.{1}^2}L_2(s_{234},s_{23})
	\nonumber \\
	&\phantom{\sum_{i=4}^{n}\bigg(}
	+\frac{\NP}{6}\frac{\langle1432\rangle}{\spa{1}.{2}}L_1(s_{341},s_{41})
	+\frac{\NP}{6}\frac{\langle2341\rangle}{\spa{2}.{1}}L_1(s_{234},s_{23})
	\nonumber \\
	&\phantom{\sum_{i=4}^{n}\bigg(}
	+\frac{\beta_0}{N}\,\frac{\langle1432\rangle}{\spa{1}.{2}}L_1(s_{341},s_{41})
	+\frac{\beta_0}{N}\,\frac{\langle2341\rangle}{\spa{2}.{1}}L_1(s_{234},s_{23})
	\bigg)
	\Bigg],
	\label{eq:Cn}
\end{align}
where 
for convenience, we have introduced
\begin{equation}
\beta_0 = \frac{11N-2\NF}{3}, \qquad\NP= 2\left(1-\frac{\NF}{N}\right),\qquad
L_k(s,t)=\frac{\log(s/t)}{(s-t)^k}
\end{equation}
and
\begin{align}
U_4 &= 
\sum_{i=1}^4 \left(\Tri(s_{i,i+2}) - \Tri(s_{i,i+3})\right)
-\frac{1}{2}\sum_{i=1}^4\Fom(s_{i,i+2};s_{i,i+1},s_{i+1,i+2})\nonumber\\
	&-\frac{1}{2}\sum_{i=1}^4\Ftme(s_{i,i+3},s_{i+1,i+2};s_{i,i+4},s_{i+1,i+3}).
\end{align}
The one-mass triangle $\Tri$ and box functions ${\rm F}_4$ can be found in~\cite{BDDK:uni1}.  

\section{Rational Contributions}
We calculate the rational part using the unitarity bootstrap proposed by Bern, Dixon and Kosower
\cite{Bern:bootstrap} which generalised
the tree level recursion of Britto, Cachazo and Feng \cite{Britto:rec}. 
The method relies on simple complex analysis and the factorisation properties of one-loop
amplitudes \cite{Bern:1lmultifact}.
In order to use this method it is very important to first remove all spurious
singularities from the pure cut terms. Once this is achieved the remaining rational terms can be
calculated using a recursion relation.

The spurious poles in eq. \eqref{eq:Cn} appear in the functions $L_2$ and $L_3$, which can be
removed by replacing these functions with new functions $\wh{L}_2$ and $\wh{L}_3$
\begin{align}
	L_i(s,t) = \wh{L}_i(s,t)+\frac{1}{2(s-t)^{i-1}}\left(\frac{1}{t}+\frac{1}{s}\right),
	\qquad i=2,3.
	\label{eq:newlogbasis}
\end{align}
The additional rational terms must then be subtracted off again such that $C_4 = \wh{C}_4+CR_4$ with
$CR_4$ given by:
\begin{align}
	CR_4(\phi,1^-,2^-,3^+,4^+) = C_4[L_1,L_2,L_3]-C_4[\wh{L}_1,\wh{L}_2,\wh{L}_3]
\end{align}

\begin{figure}[h]
	\begin{center}
		\psfrag{1}{\tiny$\wh{1}^-$}
		\psfrag{2}{\tiny$\wh{2}^-$}
		\psfrag{3}{\tiny$3^+$}
		\psfrag{4}{\tiny$4^+$}
		\psfrag{ip1}{\tiny$(i+1)^+$}
		\psfrag{i-1}{\tiny$(i-1)^+$}
		\psfrag{n-1}{\tiny$3^+$}
		\psfrag{n}{\tiny$4^+$}
		\psfrag{i}{\tiny$i^+$}
		\psfrag{phi}{\tiny$\phi$}
		\psfrag{p}{\tiny$+$}
		\psfrag{m}{\tiny$-$}
		\includegraphics[width=14cm]{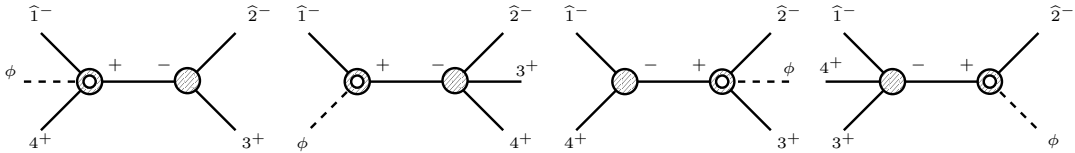}
	\end{center}
	\caption{The direct recursive diagrams contributing to $R_4(\phi,1^-,2^-,3^+,4^+)$ with a
	$|1|\la2|$ shift. }
	\label{fig:1lrec}
\end{figure}

The direct recursive terms are calculated by making a shift into complex momenta. In the case of the
$\phi$-MHV amplitudes on can avoid all boundary terms and non-factorising 3-point loop amplitudes by
choosing $\wh{p}_1 = p_1+z|2\ra[1|$ and $\wh{p}_2 = p_2-z|2\ra[1|$. This shift gives us a recursion
relation in terms of lower point $\phi$-MHV amplitudes, known finite $\phi$-amplitudes
\cite{Berger:higgsrecfinite} and QCD amplitudes leading to a rational term $R^D$. 
The recursive part of the rational contribution is defined by,
\begin{equation}
R_4^D = \sum_i\frac{A_L^{(0)}(z)R_R(z)+R_L(z)A_R^{(0)}(z)}
{P_i^2}.
\end{equation}
For the $|1|\la2|$ shift the contributing diagrams are shown in fig.~\ref{fig:1lrec}. 

The final step is to remove overlap terms
which appear due to a double counting of poles in $CR_4$. These are computed by evaluating $CR_4(z)/z$
at the poles in recursion:
\begin{equation}
	O_4 = \sum_{\alpha} \frac{CR_4(z_\alpha)P_\alpha(z_\alpha)}{P_\alpha(0)}.
	\label{eq:overlap}
\end{equation}
Collecting results for the four gluon case~\cite{Badger:phimm} and constructing the Higgs amplitude yields,
\begin{eqnarray}
R_4(H;1^-,2^-,3^+,4^+)&=&G(1,2,3,4,\A{~}{~},\B{~}{~})+G(2,1,4,3,\A{~}{~},\B{~}{~})\nonumber\\
&&+G(3,4,1,2,\B{~}{~},\A{~}{~})+G(4,3,2,1,\B{~}{~},\A{~}{~}) 
\end{eqnarray}
where
\begin{align}
G(&1,2,3,4,\A{~}{~},\B{~}{~})=\nonumber \\
&
\frac{\NP}{96\pi^2}\bigg[
2\frac{\A{2}{3}^2\B{3}{4}^3\A41}{\A34\B14\AB{3}{{41}}{3}^2}
-\frac{\A{2}{3}^2\B{3}{4}^3}{\A34\B{1}{4}^2\AB3{41}3}
+3\frac{\A12\A23\B{3}{4}^2}{\A34\B14\AB3{41}3}\nonumber \\
&-\frac{\AB{4}{{13}}{4}\AB{2}{{13}}{4}^2}{s_{341}\B{4}{1}^2\A{3}{4}^2}
+\frac{s_{412}\B34\A23}{\B41\B12\A{3}{4}^2}
+\frac{\AB{2}{{13}}{4}\A 12}{\B41\A{3}{4}^2}
-\frac14\bigg(\frac{\B34}{\B12}-\frac{\A12}{\A34}\bigg)^2
\bigg]. 
\end{align}

\section{Conclusions}
We have employed four-dimensional unitarity and 
recursion relations to compute the one-loop corrections to
a specific amplitude involving four
gluons and 
a colourless scalar - the Higgs boson.
The amplitude presented here may be useful in 
computing the gluon fusion contamination of the weak boson fusion signal for
events containing a Higgs and two jets at the LHC.

\section*{Acknowledgements}

This work was supported by Agence Nationale de la Recherche grant ANR-05-BLAN-0073-01 and SB's attendance at Moriond QCD was supported by a Marie-Curie Travel Bursary. 

\section*{References}

\providecommand{\href}[2]{#2}\begingroup\raggedright\endgroup

\end{document}